\documentclass[utf8]{jetp} 
\twocolumn 
\usepackage{amsmath}
\usepackage{graphicx}
\usepackage{amsfonts}
\usepackage{amssymb}
\usepackage{longtable}

\begin{document}

\title{Constraints of extended gravity theory in the regime of Universe accelerated extension: turnaround radius}

\thanks{Constraints of extended gravity theory ...}%

\setaffiliation1{Department of Quantum Theory and High Energy Physics, Physics Faculty, Lomonosov Moscow State University, Vorobievi Gori, 1/2, Moscow, 119234, Russia}

\setaffiliation2{Sternberg Astronomical Institute, Lomonosov Moscow State University, Universitetskii Prospekt, 13, Moscow, 119234, Russia}

\setauthor{O.}{Zenin}{1}
\setauthor{S.}{Alexeyev}{12}

\date{March 18, 2025}

\abstract{
Applying the definition of the turnaround radius and the fact that the best agreement with observational data on extragalactic scales is currently provided by general relativity with the cosmological constant we consider the behaviour of spherically symmetric solutions of the Horndeski and Dvali-Gabadadze-Poratti models on these scales. Therefore, the conditions on the turning radius together with astronomical data for galaxy clusters allow us to place additional constraints on the parameters of extended gravity theories.
}

\maketitle

\section{Introduction}\label{sec1}

At present the fully accepted and experimentally verified theory of gravity is the general theory of relativity (GR). It explains such observed phenomena as the dynamics of close binary systems, gravitational waves \cite{ligo1}, direct images of black holes (BH) \cite{EventHorizonTelescope:2022apq,EventHorizonTelescope:2019pgp}. So to take into account the accelerated expansion of the Universe it is necessary to introduce an additional term into the Lagrangian of the theory: the cosmological constant. To explain the nature and origin of this cosmological constant different extended theories of gravity are being developed: scalar fields of various nature, scalar-tensor gravity, higher-order curvature corrections, models of the type ``brane world''. Thus, there is a demand to test the behaviour of these extended theories on various space-time scales \cite{Alexeyev:2022_2} including large (extra-galactic) distances. Therefore one can consider a hypothetical surface far from clusters body on which the action of internal attractive forces is balanced by the accelerated expansion of the Universe: the turnaround radius \cite{chernin}. Since the value of this turnaround radius can be estimated from astronomical data (based on the $\Lambda$CDM theory) on the one hand, and calculated on the other, the conditions on the turning radius (the equation and inequality) can be used to estimate the behaviour of extended gravity theories on the scale of clusters (where the contribution of the accelerated expansion of the Universe is already significant). Earlier this idea was tested on the Starobinsky model with a vanishing cosmological constant \cite{Alexeyev:2017_1}, then the consideration was extended to metrics in which $g_{11} \neq -g_{00}^{-1}$ \cite{Alexeyev:2020_1}. Subsequently, the consideration was extended to other theories \cite{Nemtinova1} used in modelling black hole shadows in \cite{Alexeyev:2025_1}.
 
The structure of the paper is as follows: Section 2 is devoted to the turnaround radius finding, Section 3 is devoted to a discussion of the behaviour of different versions of the Horndeski and Dvali-Gabadze-Poratti (DGP) models on extragalactic scales, and Section 4 contains a discussion and conclusions.

\section{Conditions on turnaround radius}

At \cite{Nemtinova1} we discuss how to obtain the gravitational potential at a big distance from the centre for an arbitrary spherically symmetric metric. So:
\begin{equation}\label{metric} 
ds^2 =  - A(r) dt^2 + B(r) dr^2 + r^2\left( d\theta^2+\sin^2{\theta} d\phi^2\right),
\end{equation}
where $|B(r)| \neq |A(r)|^{-1}$, the potential $\phi$ is:
\begin{eqnarray}\label{phi0}
\phi & = & \frac{c^2}{2}B^{-1}(r)(1-A^{-1}(r)),
\end{eqnarray}

The condition on the turnaround radius (a hypothetical surface located far from the cluster centre where the action of the internal gravitational forces is balanced by the accelerated expansion of the Universe) is the maximum of the gravitational potential, therefore, 
\begin{equation}\label{eq2}
\frac{d\phi}{dr}=0, \qquad \frac{d^2\phi}{dr^2} < 0.    
\end{equation}

Rewrite the metric (\ref{metric}) for $\Lambda$CDM in SI units:
\begin{eqnarray}\label{B_metric_2}
A(r) & = & 1-\frac{2GM}{c^2 r}-\frac{\Lambda}{3}r^2,\nonumber \\   
B^{-1}(r) & = & A(r),
\end{eqnarray}
where $G=6.67 \times 10^{-11}\frac{м^3}{с^2 кг}$ is gravitational constant, $c=3 \times 10^8 \frac{м}{с}$ is speed of light in vacuum, $\Lambda=1.1 \times 10^{-52} м^{-2}$ is modern value of cosmological constant from $\Lambda CDM$ \cite{barrow1,pdg}. The derivative of the gravitational potential vanishing at the turnaround radius takes the form:
\begin{eqnarray}\label{dphi}
\frac{d\phi}{dr} & = & \frac{GM}{r^2}-\frac{\Lambda c^2 r}{3}=0.
\end{eqnarray}
From \eqref{dphi} we obtain an expression for the turnaround radius $r_t$ in GR:
\begin{eqnarray}\label{rt}
r_t & = & \Biggl( \frac{3GM}{\Lambda c^2} \Biggr)^{1/3}.
\end{eqnarray}
Thus when analysing the behaviour of extended theories of gravity on extragalactic scales we consider $r_t$ with the cosmological constant $\Lambda$ as a known quantities. 

\section{Constraints on Horndeski theory and DGP model}

The best way to account for the accelerated expansion of the Universe on extragalactic scales is to include an additional term i.e. the cosmological constant in the Lagrangian. Most modern extended theories of gravity proceed in a similar way, including such additional constant. Only a few models consider more general constructions that generate the de Sitter asymptotic do not explicitly including the $\Lambda$-term. There are for example the Horndeski theories \cite{hor1,hor2} or the ``braneworld'' models \cite{smolyakov} (a striking example of which is the DGP model \cite{dgp}).   

\subsection{Horndeski theory, case 1}

Consider the metric for Horndeski theory proposed by the formula (29) from \cite{babichev2023}. Neglecting rapidly decreasing terms, we represent it as: 
\begin{eqnarray} 
A(r) & = & B^{-1} (r) = \nonumber \\ & = & 1 - \frac{2GM}{c^2 r} - \frac{2\alpha_4}{r^2} - \frac{8\alpha_5 \eta}{5r^3} - \frac{\Lambda}{3}r^2,
\end{eqnarray}
where $\alpha_5$, $\alpha_4$ and $\eta$ are theory parameters describing the renormalization of mass and cosmological constant. The equation on turnaround radius takes the form:
\begin{equation}\label{dphi_horn} 
\frac{d\phi}{dr} =  \frac{GM}{r_t^2} + \frac{2\alpha_4 c^2}{r_t^3} + \frac{12\alpha_5 \eta c^2}{5r_t^4} - \frac{\Lambda}{3}c^2 r_t = 0.
\end{equation}
Add the condition on second derivative:
\begin{align}\label{d2phi_horn} 
\frac{d^2\phi}{dr^2} =  -\frac{2GM}{r_t^3} - \frac{6\alpha_4 c^2}{r_t^4} - \frac{48\alpha_5 \eta c^2}{5r_t^5} - \frac{\Lambda}{3}c^2 <0.
\end{align}
The relations \eqref{dphi_horn} and \eqref{d2phi_horn} are not independent so the final formula looks as:
\begin{align}\label{horn+} 
3GMr_t^2+8\alpha_4 c^2 r_t+12\alpha_5 \eta c^2 >0.
\end{align}
We obtain new estimates for these quantities. Taking into account the results of gravitational-wave astronomy (event GW200115) it was shown that (formula (39) from \cite{babichev2023}):
\begin{equation*}
|\alpha_5 \eta| \leq 2070^{565}_{659} km^3.      
\end{equation*}
Combining this constraint with \eqref{horn+}, one gets a limit on the values  $\alpha_4$:
\begin{align}\label{alpha_4} 
\alpha_4 > -\frac{3GMr_t}{8c^2} - \frac{3\alpha_5 \eta}{2r_t}.
\end{align}
Using the dependence of the turnaround radius upon the mass and the constraints on $\alpha_5 \eta$ one can estimate the numerical value of $\alpha_4$. Take the GR expression \eqref{rt} considering the lower ($M=10^{11}M_\odot$) and upper ($M=10^{15}M_\odot$) limits of the mass values:
\begin{align}\label{alpha_4_M} 
\alpha_4 (M=10^{11}M_\odot) > -8.8 \times 10^{35}, \nonumber \\
\alpha_4 (M=10^{15}M_\odot) > -1.9 \times 10^{41}.
\end{align}
Apply the formula (42) from \cite{babichev2023} i.e. the expression for the background scalar field in the form:
\begin{equation*}
\phi_0 = \ln{\Biggl( -\frac{\alpha_4}{\alpha_5}\Biggr)}. 
\end{equation*}
As a zero approximation we apply the estimate obtained by analysing the role of the Horndeski model in the binary pulsar radiation and reduced to the field of the Brans-Dicke model \cite{mnras19}, formula (92):
\begin{equation*}
\phi_0 < 0.00004. 
\end{equation*}
So knowing constraints on $\alpha_5 \eta$ и $\alpha_4$ one can extract $\eta$ so:
\begin{align}\label{eta} 
\eta (M=10^{11}M_\odot) < 2.26\times 10^{-24}, \nonumber \\
\eta (M=10^{15}M_\odot) < 1.06\times 10^{-29}.
\end{align}
Thus by combining an analysis based on the consideration of gravitational wave astronomy with additions on extragalactic scales it becomes possible to specify the permissible ranges of the theory parameters which were previously simply free.  
\subsection{Horndeski theory, case 2}

Consider another metric of Horndeski theory: metric (3) from \cite{horn3} (This metric was also discussed in \cite{horn4}):
\begin{eqnarray} 
A(r) & = & B^{-1} (r) = 1+\frac{r^2}{2\alpha}\Biggl( 1-\sqrt{1+\frac{8\alpha GM}{c^2r^3}}\Biggr),
\label{sh}
\end{eqnarray}
where $\alpha$ is theory parameter. The derivative of gravitational potential is:
\begin{eqnarray}\label{dphi_sh} 
\frac{d\phi}{dr} & = & \frac{c^2 r}{2\alpha}-\frac{c^2r}{2\alpha}\sqrt{1+\frac{8\alpha GM}{c^2r^3}} - \nonumber \\
& - & \frac{3GM}{r^2\sqrt{1+\frac{8\alpha GM}{c^2r^3}}} \approx \nonumber \\
& \approx & -\frac{5GM}{r^2}+\frac{12\alpha G^2M^2}{c^2r^5}=0.
\end{eqnarray}
Second derivative:
\begin{equation}\label{d2phi_dgp} 
\frac{d^2\phi}{dr^2}\approx \frac{10GM}{r^3}-\frac{60\alpha G^2M^2}{c^2r^6}<0.
\end{equation}
As a result the expression for the turnaround radius is obtained:
\begin{equation}\label{rt_sh} 
r_t=\Biggl( \frac{12\alpha GM}{5c^2}\Biggr) ^{1/3}. 
\end{equation}
From the expression for the second derivative it is seen that the condition of the turnaround radius is satisfied for any value of $\alpha$ (we obtain the identical expression $\frac{12}{5}<6$). In \cite{horn4} this restriction was more stringent: $\alpha \leq M$ thus in this case the analysis of the behaviour on extragalactic scales only confirms the constraints put earlier. 

\subsection{DGP model}

Consider the DGP theory and the metric (37) (the second branch) from \cite{dgp2}:
\begin{eqnarray} 
A(r) & = & B^{-1} (r) = 1 - \frac{2GM}{c^2r} -\frac{r^2}{2r_{c}^2}+\frac{q^2}{r^2}-\frac{q^4r_{c}^2}{5r^3}. \label{dgp}
\end{eqnarray}
Here $r_c$ is the ratio of four- and five-dimensional masses: $r_c=M^2_{(4)}/2M^3_{(5)}$, $q$ is tidal charge, characterizing the contribution from extra dimensions. We construct the gravitational potential equating its first derivative to zero:
\begin{equation}\label{dphi_dgp} 
\frac{d\phi}{dr} =  \frac{GM}{r^2}-\frac{c^2r}{2r_{c}^2}-\frac{c^2q^2}{r^3}+\frac{3c^2q^4r_c^2}{10r^4} =0.
\end{equation}
The second derivative is:
\begin{equation}\label{d2phi_dgp} 
\frac{d^2\phi}{dr^2} =  -\frac{2GM}{r^3}-\frac{c^2}{2r_{c}^2}+\frac{3c^2q^2}{r^4}-\frac{6c^2q^4r_c^2}{5r^5} <0.
\end{equation}
Considering the turnaround radius $r_t$ and the mass $M$ as known values we obtain an additional constraint on the theory:
\begin{align}\label{dgp_rt} 
\frac{3}{2}c^2r_c^2q^4 - 5c^2r_tq^2 + 3GMr_t^2 > 0.
\end{align}
So we obtained the relation between $q$ and $r_c$. Substituting the values for the turnaround radius from GR \eqref{rt}, we obtain a bi-quadratic equation with respect to $q$. From the solution existence condition and substituting the boundary value $\acute{q} < 10^{-17}$ we obtain the constraint: $r_c < 3.3 \times 10^{14}$ which means that at large distances together with the de Sitter asymptotes the divergent regime $r_c\to \infty$ is realized as it was predicted in \cite{dgp2}. Note that the obtained value has the same order of magnitude as the Weinstein radius determined by formula (40) \cite{dgp2}. Thus here our analysis confirms the constraints on the parameters of the theory put earlier.

\section{Discussion and Conclusions}

We propose an approach that allows us to impose additional constraints on the theory parameters during testing the predictions of the extended theory of gravity on extragalactic scales. Due to the insufficient astronomical data we have to use formulas for GR with the cosmological constant which naturally reduces the accuracy of the estimates. Thus, for the Horndeski model in the first version \cite{babichev2023} we managed to obtain estimates for a parameter that was previously considered to be free and estimated only in combination with others, for the Horndeski model in the second version \cite{horn3} we managed to confirm the parameter values \cite{horn3} obtained earlier and for the DGP model \cite{dgp2} we show the implementation of a scenario in which the world is effectively multidimensional at large distances (beyond the Weinstein radius). The cosmological constant is not explicitly included in the models considered, but on extragalactic scales the solution is reduced to be de Sitter asymptotically. The proposed method in a number of cases allows to impose additional restrictions on theories making possible to use it in the future in a system of tests for extended theories of gravity (in the development of the parameterized post-Newtonian formalism) \cite{Alexeyev:2022_2} on various energy and distance scales to impose new restrictions on the parameters of the theory. 

\section{Acknowledgments}

The work was supported via grant from the Russian Science Foundation No. 23-22-00073.


\begin{thebibliography}{100}

\bibitem{ligo1}
LIGO Scientific and VIRGO Collaborations: R. Poggiani and others, PoS MULTIF2023 021 (2024).

\bibitem{EventHorizonTelescope:2022apq}
The Event Horizon Telescope Collaboration: K. Akiyama, A. Alberdi, W. Alef and others, Astrophys. J. Lett. {\bf 930}, L13 (2022). 

\bibitem{EventHorizonTelescope:2019pgp}
The Event Horizon Telescope Collaboration: K. Akiyama, A. Alberdi, W. Alef and others, Astrophys. J. Lett. {\bf 875}, L5, (2019).

\bibitem{Alexeyev:2022_2}
S. Alexeyev and V. Prokopov, Universe {\bf 8}, 283 (2022).

\bibitem{chernin}
A. D. Chernin, N. V. Emelyanov and I. D. Karachentsev, Mon.Not.Royal Astron.Soc. {\bf 449}, 2069 (2015).

\bibitem{Alexeyev:2017_1}
S.Alexeyev, B.Latosh, V.Echeistov, JETP {\bf 125}, 1083 (2017).

\bibitem{Alexeyev:2020_1}
S. Alexeyev and K. Kovalkov, Int. J. Mod. Phys. A {\bf 35}, 204057 (2020).

\bibitem{Nemtinova1}
A.Nemtinova, S.Alexeyev, {\it Constraints on gravity models at galaxy cluster scales} Physics pf the Universe: Proceedings of 50 student science workshop, Yekaterinburg, January 30 - February 3, 2023, p.248, Yekaterinburg, Ural Federal University (2023), ISBN 978-5-7996-3700-2 (in Russian).

\bibitem{Alexeyev:2025_1}
S.Alexeyev, O.Zenin and A.Baiderin, JETP, {\bf 167}, 477 (2025).

\bibitem{barrow1}
J.D. Barrow and D.J. Shaw, Int. J. Mod. Phys. D {\bf 20}, 2875 (2011).

\bibitem{pdg}
Particle Data Group \underline{https://pdg.lbl.gov/}

\bibitem{hor1}
G. Horndeski, Int. J. Theor. Phys. {\bf 10}, 363 (1974).

\bibitem{hor2} 
T. Kobayashi, Rept. Prog. Phys. {\bf 82}, 086901 (2019).

\bibitem{smolyakov}
E.Boos, V.Bunichev, I.Volobuev, V.Smolyakov, Phys.Part.Nucl. {\bf 43}, 1 (2012).

\bibitem{dgp}
G.R. Dvali, G. Gabadadze and M. Porrati, Phys.Lett.B {\bf 485}, 208 (2000).

\bibitem{babichev2023}
E. Babichev, C. Charmousis, M. Hassaine and N. Lecoeur, Phys.Rev.D {\bf 108}, 024019 (2023).

\bibitem{horn3}
A. Bakopoulos, C. Charmousis, P. Kanti, N. Lecoeur and T. Nakas, Phys.Rev.D {\bf 109}, 024032 (2024).

\bibitem{horn4}
H. Huang, J. Kunz and D. Mitra, JCAP {\bf 05}, 07 (2024).

\bibitem{dgp2}
R. Gannouji, Eur.Phys. C {\bf 78}, 318 (2018).

\bibitem{mnras19}
P. Dyadina, N. Avdeev and S. Alexeyev, Mon. Not. Royal Astr. Soc. {\bf 483}, 947 (2019).

\end{thebibliography}
\end{document}